**Coordinate-free description of corrugated flames with realistic density drop at the front**

Vitaly Bychkov,[1] Maxim Zaytsev,[1,2] and V'yacheslav Akkerman,[1,2]

[1] *Institute of Physics, Umeå University, S-901 87 Umeå, Sweden*

[2] *Moscow Institute of Physics and Technology, 141 700, Dolgoprudny, Russia*

**Abstract**

The complete set of hydrodynamic equations for a corrugated flame front is reduced to a system of coordinate-free equations at the front using the fact that vorticity effects remain relatively weak even for corrugated flames. It is demonstrated how small but finite flame thickness may be taken into account in the equations. Similar equations are obtained for turbulent burning in the flamelet regime. The equations for a turbulent corrugated flame are consistent with the Taylor hypothesis of "stationary" external turbulence.

PACs: 82.40 Py, 47.20 k, 47.27 i

**Communication address:**

V. Bychkov,

Inst. of Physics, Umeå University

S-901 87, Umeå, Sweden

tel. (46 90) 786 79 32     fax:  (46 90) 786 66 73

e-mail:  vitaliy.bychkov@physics.umu.se

## 1. Introduction

One of the most difficult problems of premixed combustion modelling is huge difference in length scales involved in the burning process. While the characteristic length scale of the hydrodynamic flow varies from 5 - 10 cm (car engines) to several meters (turbine combustors), the flame thickness is typically much smaller $L_f = \left(10^{-4} - 10^{-3}\right)$cm, and the zone of active reactions is even thinner $\approx 0.1 L_f$ (Williams, 1985). No computer can resolve all these length scales at present, and therefore one of the main tasks of combustion science is to create a reliable model of burning, both turbulent and laminar. The problem may be simplified considerably, if one manages to reduce the whole system of combustion equations to a single equation for the flame front position. This is possible, for example, for a turbulent flame in the artificial case of zero density variations across the flame front, when the ratio of the fuel mixture to burnt gas density is unity $\Theta \equiv \rho_f / \rho_b = 1$ (Kerstein et al., 1988; Yakhot, 1988). In that case a flame front propagates in a prescribed external turbulent flow without affecting the flow. However, in reality the expansion factor $\Theta$ is rather large $\Theta = 5 - 10$, flame interacts strongly with the flow, which leads to many additional phenomena such as the Darrieus-Landau (DL) instability (Williams, 1985; Zeldovich et al., 1985). As a matter of fact, attempts to simplify the whole system of combustion equations for a laminar flame with the expansion factor $\Theta$ different from unity have been usually coupled to studies of the DL instability, see, (Bychkov and Liberman, 2000) as one of the latest reviews on the subject. The basic idea of the simplifications is the following. A flame front is typically very thin in comparison with the hydrodynamic length scales, and it may be considered as a geometrical surface of zero thickness separating fuel mixture and products of burning. In that case the solution to the combustion equations consists of three steps: 1) we have to solve equations of ideal hydrodynamics in the fuel mixture ahead of the flame front; 2) we have to solve equations of ideal hydrodynamics in the burnt matter behind the flame front; 3) we have to match the obtained solutions at the flame front with the help of conservation laws. If we succeed in these three steps of solution, then we find a single nonlinear equation of the front dynamics (or a set of equations), which contains only values and derivatives at the flame front but not in the bulk of the gas flow. In the case of a flame front of zero thickness the matching conditions are specified by the conservation laws at a hydrodynamic discontinuity surface (Landau and Lifshitz, 1989) plus the condition of constant velocity $U_f$ of flame propagation with respect to the fuel mixture. However, the growth rate of the DL instability is not limited for a flame front of zero thickness, the evolution problem of an infinitely thin flame cannot be specified self-consistently, and we have to consider finite flame thickness (Williams, 1985; Pelce and Clavin, 1982). Small but finite flame thickness may be taken into account as a parameter in the conservation laws at the flame front, which have been obtained in (Clavin and Williams, 1982; Pelce and Clavin, 1982) for the linear approximation and in (Matalon and Matkowsky, 1982) for a strongly curved flame in the nonlinear regime. Respective expression for the flame propagation velocity depending on the flame stretch has been derived in (Matalon and Matkowsky, 1982, Clavin and Joulin, 1983). The conservation laws specify the step 3 in the above recipe. Solution to the hydrodynamic equations in the fuel mixture ahead of the flame front (step 1) is rather simple in the case of laminar burning, since ahead of the laminar flame the flow is potential no matter how corrugated is the front. The last and the most difficult part of the algorithm in the laminar case is to solve the hydrodynamic equations in the downstream flow behind the flame (step 2). Indeed, a curved flame front generates vorticity in the flow, which makes the flow essentially nonlinear (Zeldovich et al., 1985; Bychkov and Liberman, 2000). If we consider turbulent burning, then vorticity is non-zero both ahead and behind the flame front, which makes the problem even more difficult. Therefore, trying to develop the model of a turbulent thin flame front, we have to solve first a similar laminar problem.

Up to now the problem of laminar corrugated flame dynamics has been reduced to a single equation for the flame front position only under simplifying assumptions. Sivashinsky has derived an equation of this kind in the artificial limit of small thermal expansion $\Theta - 1 << 1$ assuming also weak nonlinearity of the flame front, i.e. that a flame shape differs slightly from an ideally planar or ideally spherical front (Sivashinsky, 1977). In the case of an arbitrary thermal expansion including realistically large expansion factors $\Theta = 5 - 10$ an equation of flame front evolution has been derived in (Bychkov, 1998) using the same assumption of weak nonlinearity. The obtained equation described successfully velocity of 2D curved stationary flames (Bychkov, 1998, Bychkov et al., 1996) and stability limits of the curved flames (Bychkov et al. 1999; Travnikov et al., 2000). Unfortunately, the assumption of weak nonlinearity has rather limited number of applications, for example, it cannot be applied to strongly nonlinear fractal flames expected at large hydrodynamic length scales (Bychkov and Liberman, 2000; Gostintsev et al., 1988; Bradley et al., 2001; Aldredge and Zuo, 2001). In order to describe fractal flames we have to derive an equation (or a set of equations) at a flame front in a coordinate-free rotationally invariant form without any restriction on the nonlinear terms. Such an equation has been derived by Frankel (Frankel, 1990) in the artificial limit of small thermal expansion $\Theta - 1 \ll 1$ similar to the Sivashinsky equation (the Frankel equation may be reduced to the Sivashinsky equation for a weakly nonlinear front). The Frankel equation has been widely used (Blinnikov and Sasorov, 1996; Denet, 1997; Peters et al. 2000; Denet, 2002) because it describes qualitative behaviour of strongly corrugated flames. However, the limit of small thermal expansion $\Theta - 1 \ll 1$ adopted by Frankel is too far from parameters of realistic flames with the expansion factors $\Theta = 5 - 10$ and cannot be utilized for a quantitative analysis. Therefore, what is needed is an equation (or a set of equations) at a discontinuous flame front written in a coordinate-free form similar to the Frankel equation, but taking into account realistically large thermal expansion of burning matter.

In the present paper we reduce the complete set of hydrodynamic equations for a corrugated flame front to a system of coordinate-free equations at the front using the fact that vorticity effects remain relatively weak even for a fractal flame. We demonstrate how small but finite flame thickness may be taken into account in the equations. We show that similar equations may be obtained for turbulent burning in the flamelet regime. The equations obtained for a turbulent corrugated flame are consistent with the Taylor hypothesis of "stationary" external turbulence.

## 2. Basic equations for an infinitely thin flame front

We start with the DL approximation of an infinitely thin flame front. Suppose that gas dynamics of burning is characterised by a length scale $R$, which may be radius of a tube, width of a channel, radius of a spherical burning chamber, etc. We introduce the dimensionless velocity of the flow scaled by the velocity of a planar flame front $\mathbf{u} = \mathbf{v}/U_f$, together with the scaled coordinates $\mathbf{r} = \mathbf{x}/R$, time $\tau = U_f t/R$ and pressure $\Pi = (P - P_f)/\rho_f U_f^2$, where $\rho_f$ and $P_f$ are density and pressure in the fuel mixture far ahead of the flame front. Within the framework of the DL approximation the flame is treated as a discontinuity surface of zero thickness propagating at a constant velocity $U_f$ relative to the fuel mixture. The flow is assumed to be incompressible and inviscid, and the hydrodynamic equations upstream and downstream of a corrugated flame take the form

$$\nabla \cdot \mathbf{u} = 0, \qquad (1)$$

$$\frac{\partial \mathbf{u}}{\partial \tau} + (\mathbf{u} \cdot \nabla)\mathbf{u} = -\vartheta \nabla \Pi, \tag{2}$$

where $\vartheta = 1$ in the fuel mixture and $\vartheta = \Theta$ in the burnt gas. Let the flame front be described by the function

$$F(\mathbf{r}, \tau) = 0. \tag{3}$$

We choose $F(\mathbf{r}, \tau) < 0$ in the fuel mixture and $F(\mathbf{r}, \tau) > 0$ in the products of burning, so that the normal unit vector $\mathbf{n} = \nabla F / |\nabla F|$ points to the products, see Fig. 1. The geometrical surface $F(\mathbf{r}, \tau) = 0$ corresponding to the flame front propagates normally to itself in the outward direction with velocity $-\mathbf{n} V_s$

$$V_s = |\nabla F|^{-1} \frac{\partial F}{\partial \tau}. \tag{4}$$

Jump conditions across an infinitely thin front are (Landau and Lifshitz, 1989)

$$u_{n+} + V_s = \Theta(u_{n-} + V_s), \tag{5}$$

$$\mathbf{u}_{t+} = \mathbf{u}_{t-}, \tag{6}$$

$$\Pi_+ + \frac{1}{\Theta}(u_{n+} + V_s)^2 = \Pi_- + (u_{n-} + V_s)^2. \tag{7}$$

Here labels "-" and "+" correspond to the positions just ahead and just behind the flame front, while the labels $n$ and $t$ stand for the normal and tangential directions. One more condition is that the flame front propagates at a fixed speed with respect to the fuel mixture

$$u_{n-} + V_s = 1. \tag{8}$$

Using Eq. (8) one can rewrite Eqs. (5) – (7) in the form

$$u_{n+} = u_{n-} + \Theta - 1, \tag{9}$$

$$\mathbf{u}_+ = \mathbf{u}_- + (\Theta - 1)\mathbf{n}, \tag{10}$$

$$\Pi_+ = \Pi_- + 1 - \Theta. \tag{11}$$

Equations (1), (2) in the upstream and downstream flow together with the conditions at the flame surface (8) - (11) describe dynamics of an infinitely thin flame front.

Though the system (1), (2), (8) - (11) looks rather simple, in reality these equations are very difficult even for a numerical study, because they require solution in the bulk of a gas both ahead and behind the flame front. The purpose of the present paper is to reduce the whole system (1), (2), (8) - (11) to a set of equations at the flame front, which is possible only under certain simplifying assumptions specified below.

## 3. The approximation of small vorticity in the flow of burned matter behind a curved flame front

In this section we discuss simplifying assumptions, which can be used in describing dynamics of corrugated flames. We start with propagation of a curved laminar flame in a tube with boundary conditions of adiabatic walls and ideal slip at the walls. The advantage of such geometry is that shape of a curved flame front in a tube may be limited by only one stationary cell, which is easier for analysis than the fractal flame shape consisting of a large number of cells of different sizes imposed on each other (Gostintsev et al., 1988; Bychkov and Liberman, 2000). According to the numerical simulations (Bychkov et al., 1996; Kadowaki, 1999; Travnikov et al. 2000) a flame cell may be described as a cusp pointing to the burnt matter and a hump directed to the fresh fuel mixture. Taking into account the tube geometry we choose scaled variables in the form $\mathbf{r} = (\mathbf{y}, \xi)$, $\mathbf{u} = (\mathbf{w}, \upsilon)$, where $\xi$ is the coordinate axis along the walls. To be particular we take the reference frame of a planar flame front, in which the fuel mixture at infinity moves towards the flame with velocity $\mathbf{u} = \mathbf{e}_{\xi}$. As we pointed out in the Introduction, we are interested first of all in the flow of the burning products. If the flame front is planar, then velocity and pressure behind the flame are uniform $\mathbf{u} = \Theta \mathbf{e}_{\xi}$, $\Pi = -\Theta + 1$ (Bychkov 1998), but in the case of a curved flame the flow is different from the uniform one. Velocity deviations from the planar flow in the products of burning $\tilde{\mathbf{u}} = \mathbf{u} - \Theta \mathbf{e}_{\xi}$ satisfy the equation

$$\frac{\partial \tilde{\mathbf{u}}}{\partial \tau} + \Theta \frac{\partial \tilde{\mathbf{u}}}{\partial \xi} + (\tilde{\mathbf{u}} \cdot \nabla) \tilde{\mathbf{u}} + \Theta \nabla \Pi = 0. \tag{12}$$

Equation (12) may be also presented in the form

$$\frac{\partial \tilde{\mathbf{u}}}{\partial \tau} + \Theta \frac{\partial \tilde{\mathbf{u}}}{\partial \xi} - \tilde{\mathbf{u}} \times \omega + \nabla \left( \frac{\tilde{u}^2}{2} + \Theta \Pi \right) = 0, \tag{13}$$

or

$$\left( \frac{\partial}{\partial \tau} + \Theta \frac{\partial}{\partial \xi} \right) \nabla \times \tilde{\mathbf{u}} - \nabla \times (\tilde{\mathbf{u}} \times \omega) = 0, \tag{14}$$

where $\omega = \nabla \times \mathbf{u}$ is vorticity (Landau and Lifshitz, 1989). Characteristic flow behind a curved flame cell obtained in numerical simulations (Bychkov et al., 1996) is shown in Fig. 2. An important point about Fig. 2 is that, in spite of the curved flame shape, the main velocity component in the flow of burned matter is determined by the uniform drift velocity $\Theta \mathbf{e}_{\xi}$, while the velocity deviations from the uniform flow are rather weak $\tilde{u} << \Theta$. According to (Bychkov, 1998) the relative role of the velocity deviations may be estimated as $\tilde{u} / \Theta \propto f / \lambda$, where $f$ and $\lambda$ are the characteristic amplitude and width of the flame cell. Direct numerical simulations of flame dynamics show that the amplitude of one cell of a curved flame is rather small $f / \lambda = 0.25 - 0.35$ even for realistically large thermal expansion $\Theta = 5 - 10$ both for 2D and 3D geometry (Bychkov et al., 1996; Kadowaki 1999; Travnikov et al., 2000). Actually, this small curvature of the flame front is the reason why the limit of weak nonlinearity is so successful in describing dynamics of a single flame cell (Bychkov, 1998; Bychkov et al., 1999). The relative role of vorticity in the flow of burning products may be specified by the combination $\omega \lambda / \Theta$. It was shown in (Bychkov, 1998), that vorticity effects behind a curved flame front are as small as $\omega \lambda / \Theta \propto \left( f / \lambda \right)^2$. The secondary role of vorticity in the DL instability has been discussed for a long time (Sivashinsky, 1977; Frankel, 1990; Peters et al. 2000). Particularly, it has been demonstrated in (Sivashinsky, 1977) that vorticity is just a by-product of the DL instability, which may be neglected completely for small thermal expansion $\Theta - 1 << 1$. The assumption of zero vorticity in the burnt matter was the basis of the Frankel equation (Frankel, 1990). However, neglecting vorticity in the burnt gas for realistically large expansion factors $\Theta = 5 - 10$ one comes to an incorrect dispersion relation at the linear stage of the DL instability. Therefore in the present paper we take vorticity into account in the linear approximation. We propose to neglect the last term of Eq. (14), namely $\tilde{\mathbf{u}} \times \omega$, which stands for the nonlinear coupling between the flame-generated vorticity and the small deviations of the flow velocity from the drift velocity. According to the above estimates the role of this nonlinear term is as small as $\left| \tilde{\mathbf{u}} \times \omega \right| \lambda / \Theta^2 \propto \left( f / \lambda \right)^3 = 0.01 - 0.04$, which is definitely beyond the computational accuracy of the direct numerical simulations (Bychkov et al., 1996; Kadowaki, 1999; Travnikov et al., 2000). Using such an approximation (below we will call it the approximation of small vorticity) we reproduce correctly the linear dispersion relation of the DL instability for any expansion factor of the burning matter. Indeed, in that case Eq. (14) becomes

$$\left(\frac{\partial}{\partial\tau}+\Theta\frac{\partial}{\partial\xi}\right)\nabla\times\tilde{\mathbf{u}}=0\,, \tag{15}$$

with the solutions corresponding to the mode of vorticity drift

$$\left(\frac{\partial}{\partial\tau}+\Theta\frac{\partial}{\partial\xi}\right)\tilde{\mathbf{u}}_v=0\,, \tag{16}$$

and to the potential mode $\nabla\times\tilde{\mathbf{u}}_p=0$, $\nabla^2\tilde{\mathbf{u}}_p=0$ with

$$\tilde{\upsilon}_p=\frac{1}{4\pi^2}\int\tilde{\upsilon}_{pk}\exp(-k\xi+i\mathbf{k}\cdot\mathbf{y})\,d^2k\,, \tag{17}$$

$$\tilde{\mathbf{w}}_p=-\hat{\Phi}^{-1}\nabla_{\perp}\tilde{\upsilon}_p\,. \tag{18}$$

Here $\hat{\Phi}$ stands for the DL operator

$$\hat{\Phi}F=\frac{1}{4\pi^2}\int kF_k\exp(i\mathbf{k}\cdot\mathbf{y})\,d^2k\,, \tag{19}$$

$F_k$ is the Fourier transform of $F$ and $\nabla_{\perp}$ corresponds to the $\nabla$-component in the plane perpendicular to the walls. One more important property of the potential mode is

$$\left(\frac{\partial}{\partial\tau}+\Theta\frac{\partial}{\partial\xi}\right)\tilde{\mathbf{u}}_p+\nabla\left(\frac{\tilde{u}^2}{2}+\Theta\Pi\right)=0\,, \tag{20}$$

which means that the dynamical pressure $\tilde{u}^2/2+\Theta\Pi$ satisfies the Laplace equation in the burnt matter

$$\nabla^2\left(\frac{\tilde{u}^2}{2}+\Theta\Pi\right)=0\,. \tag{21}$$

Obviously, the dynamical pressure obeys the Laplace equation in the fuel mixture ahead of the flame front too, because the flow in the fuel mixture is potential. The solution to Eq. (15) in the fuel mixture is

$$\upsilon=1+\tilde{\upsilon}=1+\frac{1}{4\pi^2}\int\tilde{\upsilon}_k\exp(k\xi+i\mathbf{k}\cdot\mathbf{y})\,d^2k\,, \tag{22}$$

$$\mathbf{w}=\hat{\Phi}^{-1}\nabla_{\perp}\tilde{\upsilon}\,, \tag{23}$$

$$\Pi=\frac{1}{2}-\hat{\Phi}^{-1}\frac{\partial\upsilon}{\partial\tau}-\frac{\tilde{u}^2}{2}\,. \tag{24}$$

Taking matching conditions (10), (11) at a flame front $F(\mathbf{r},\tau)=\xi-f(\mathbf{y},\tau)=0$ in the linear approximation

$$\tilde{\upsilon}_+ = \tilde{\upsilon}_-, \qquad \mathbf{w}_+ = \mathbf{w}_- - (\Theta - 1)\nabla_\perp f, \qquad \tilde{\upsilon}_- = \frac{\partial f}{\partial \tau}, \qquad (25)$$

and substituting the modes Eqs. (16) − (18), (22) − (24) we come to the equation for the flame front perturbations

$$(\Theta + 1)\frac{\partial^2 f}{\partial \tau^2} + 2\Theta\hat{\Phi}\frac{\partial f}{\partial \tau} - \Theta(\Theta - 1)\hat{\Phi}^2 f = 0, \qquad (26)$$

which reproduces correctly the DL dispersion relation for any thermal expansion including the realistically large expansion factors $\Theta = 5 - 10$ (Landau and Lifshitz, 1989). It is important that we do not neglect the mode of vorticity drift $\mathbf{u}_v$ in the burnt matter behind a corrugated flame front, but instead we neglect only the nonlinear coupling between the potential mode and the vorticity mode. Neglecting vorticity completely similar to (Sivashinsky, 1977; Frankel, 1990; Peters et al., 2000) we would come to another dispersion relation

$$\frac{\partial f}{\partial \tau} - \frac{\Theta - 1}{2}\hat{\Phi}f = 0, \qquad (27)$$

which holds only in the limit of small thermal expansion $\Theta - 1 \ll 1$. For realistically large expansion factors $\Theta = 5 - 10$ the dispersion relation (27) provides only qualitative, but not quantitative description of the DL instability. On the contrary, the present approximation of small but non-zero vorticity allows both qualitative and quantitative studies of the flame instability.

Taking into account nonlinear corrections to Eqs. (25), (26) in the limit of weak nonlinearity $(\nabla_\perp f)^2 \ll 1$ we can demonstrate that the approximation of small vorticity describes well propagation velocity of curved stationary flames and stability limits of these flames. As a matter of fact, in that case we just have to reproduce the calculations of (Bychkov, 1998; Bychkov et al., 1999), which is a long, but straightforward procedure. A much subtler question is if the approximation of small vorticity can describe properly dynamics of strongly corrugated fractal flames. A fractal structure implies self-similar properties of a flame front at different length scales (Bychkov and Liberman, 2000; Gostintsev et al. 1988; Bradley et al. 2001). Because of the self-similarity every large cell at a flame front reproduces the shape of small cells like those observed for flames propagating in relatively narrow tubes (Bychkov et al., 1996; Kadowaki, 1999; Travnikov et al., 2000). Then the characteristic ratio $f/\lambda$ remains small for any cell of the fractal cascade, and the nonlinear effects related to the flame-generated vorticity may be neglected even for strongly corrugated fractal flames.

## 4. Coordinate-free equations at a flame front of zero thickness

In the present section we derive coordinate-free equations at a flame front of zero thickness ignited at a point. Taking into account the approximation of small vorticity, in the laboratory reference frame we can write Eq. (2) in the form

$$\frac{\partial \mathbf{u}}{\partial \tau} + \nabla p = 0, \tag{28}$$

where $p = \vartheta \Pi + u^2 / 2$ is the dynamic pressure. All variables at the flame front (e.g. $\mathbf{u}_-$, $\mathbf{u}_+$, $V_s$, etc.) depend on time and on the coordinate along the front $\mathbf{r}_s$, see Fig. 1. Respective derivatives in space $\nabla_s$ coincide with the tangential derivative along the flame front $\nabla_s = \mathbf{e}_t \cdot \nabla$, but the time derivative is related to the normal derivative in space

$$\frac{\partial}{\partial \tau_s} = \frac{\partial}{\partial \tau} - V_s \frac{\partial}{\partial \mathbf{n}}, \tag{29}$$

since the flame front propagates locally with the velocity $-\mathbf{n}V_s$. The velocity field ahead of the flame front is potential $\mathbf{u} = \nabla \phi$, $\nabla^2 \phi = 0$ which leads to the Bernoulli integral

$$\frac{\partial \phi}{\partial \tau} + p = 0, \tag{30}$$

taking the form

$$\frac{\partial \phi_-}{\partial \tau_s} + V_s u_{n-} + p_- = 0 \tag{31}$$

exactly at the flame front. The velocity field behind the flame may be presented as a combination of a potential mode and a vorticity mode $\mathbf{u} = \mathbf{u}_p + \mathbf{u}_v$ satisfying the equations $\mathbf{u}_p = \nabla \phi$, $\nabla^2 \phi = 0$ and

$$\frac{\partial \mathbf{u}_v}{\partial \tau} = 0. \tag{32}$$

It is interesting to note that the vorticity mode is time-independent in the approximation of small vorticity similar to the well-known solution of the DL instability at a spherical flame propagating outwards from a point of ignition (Istratov and Librovich, 1966). Taking into account Eq. (29) the last property may be also presented as

$$\frac{\partial \mathbf{u}_v}{\partial \tau_s} = -V_s \frac{\partial \mathbf{u}_v}{\partial \mathbf{n}}. \tag{33}$$

The respective Bernoulli integral at the flame front in the burnt matter is

$$\frac{\partial \phi_+}{\partial \tau_s} + V_s u_{pn+} + p_+ = const .\tag{34}$$

Taking into account that velocity potential is defined with the accuracy of a time-dependent function (Landau and Lifshitz, 1989), we may neglect the constant in the last equation. Boundary conditions for the velocity potentials at the flame front follow from Eqs. (8), (9)

$$\frac{\partial \phi_-}{\partial \mathbf{n}} = u_{n-} = 1 - V_s ,\tag{35}$$

$$\frac{\partial \phi_+}{\partial \mathbf{n}} = u_{pn+} = \Theta - V_s - u_{vn+} .\tag{36}$$

According to the Green solution to the Laplace equation, a harmonic function $\nabla^2 \phi = 0$ at a point $\mathbf{r}$ of a domain $G$ with a surface $S$ and a normal unit vector $\mathbf{n}_{out}$ pointing outwards may be found using the boundary conditions at the surface

$$\beta \phi(\mathbf{r}) = \iint_S \left[ \frac{1}{|\mathbf{r}_s - \mathbf{r}|} \frac{\partial \phi(\mathbf{r}_s)}{\partial \mathbf{n}_{out}} + \phi(\mathbf{r}_s) \mathbf{n}_{out} \cdot \frac{\mathbf{r}_s - \mathbf{r}}{|\mathbf{r}_s - \mathbf{r}|^3} \right] dS(\mathbf{r}_s) ,\tag{37}$$

where $\beta = 4\pi$ if $\mathbf{r}$ is inside $G$, $\beta = 2\pi$ if $\mathbf{r}$ is on the surface $S$, and $\beta = 0$ if $\mathbf{r}$ is outside $G$. Then potentials $\phi_-$, $\phi_+$ at the flame front satisfy the equations

$$\phi_-(\mathbf{r}) = \frac{1}{2\pi} \iint_S \left[ \frac{1 - V_s}{|\mathbf{r}_s - \mathbf{r}|} + \phi_- \mathbf{n} \cdot \frac{\mathbf{r}_s - \mathbf{r}}{|\mathbf{r}_s - \mathbf{r}|^3} \right] dS(\mathbf{r}_s) ,\tag{38}$$

$$\phi_+(\mathbf{r}) = -\frac{1}{2\pi} \iint_S \left[ \frac{\Theta - V_s - u_{vn+}}{|\mathbf{r}_s - \mathbf{r}|} + \phi_+ \mathbf{n} \cdot \frac{\mathbf{r}_s - \mathbf{r}}{|\mathbf{r}_s - \mathbf{r}|^3} \right] dS(\mathbf{r}_s) ,\tag{39}$$

where all values under the integrals depend on $\mathbf{r}_s$ (except for $\mathbf{r}$, of course). Equations (38), (39) are integral equations, which determine velocity potentials at the flame front, if the vorticity component of the velocity field is known just behind the flame. Therefore, we have to find the relation between the potential modes and the vorticity mode at the flame front. For that purpose we write the continuity equation (1) for the vorticity mode in the form

$$\frac{\partial u_{vn+}}{\partial \mathbf{n}} + \nabla_s \cdot \mathbf{u}_{vt+} = 0 ,\tag{40}$$

and using Eq. (33) we reduce it to

$$\frac{\partial u_{vn+}}{\partial \tau_s} = V_s \nabla_s \cdot \mathbf{u}_{vt+}. \tag{41}$$

The jump condition for the tangential velocity $\mathbf{u}_{t-} = \mathbf{u}_{t+}$, Eq. (6), leads to $\mathbf{u}_{t-} - \mathbf{u}_{pt+} = \mathbf{u}_{vt+}$ and couples the tangential velocity component of the vorticity mode and the velocity potentials as

$$\nabla_s (\phi_- - \phi_+) = \mathbf{u}_{vt+}. \tag{42}$$

Then, taking into account Eq. (41) we find the desired relation between the potential modes and the normal velocity component of the vorticity mode used in Eq. (39)

$$\frac{\partial u_{vn+}}{\partial \tau_s} = V_s \nabla_s^2 (\phi_- - \phi_+). \tag{43}$$

Finally, we have to couple the potentials. For that purpose we substitute Eqs. (31), (34) into the condition of pressure jump at the flame front Eq. (11)

$$\frac{\partial}{\partial \tau_s}(\phi_+ - \Theta \phi_-) = \Theta(\Theta - 1) - \frac{u_+^2}{2} + \Theta \frac{u^2}{2} - V_s u_{n+} + V_s u_{vn+} + \Theta V_s u_{n-}. \tag{44}$$

Using Eqs. (8) – (10) we can reduce Eq. (44) to

$$\frac{\partial}{\partial \tau_s}(\phi_+ - \Theta \phi_-) = \frac{\Theta - 1}{2} u_-^2 - (\Theta - 1) V_s^2 + (\Theta - 1) V_s + V_s u_{vn+} + \frac{(\Theta - 1)^2}{2}. \tag{45}$$

Introducing the designations

$$\Psi = \frac{\phi_+ - \Theta \phi_-}{\Theta - 1}, \qquad \Omega = \frac{u_{vn+}}{\Theta - 1}, \tag{46}$$

we can rewrite the final set of equations at the flame front in the form

$$\phi_-(\mathbf{r}) = \frac{1}{2\pi} \iint_S \left[ \frac{1 - V_s}{|\mathbf{r}_s - \mathbf{r}|} + \phi_- \mathbf{n} \cdot \frac{\mathbf{r}_s - \mathbf{r}}{|\mathbf{r}_s - \mathbf{r}|^3} \right] dS(\mathbf{r}_s), \tag{47}$$

$$\Psi(\mathbf{r}) + \frac{\Theta + 1}{\Theta - 1} \phi_-(\mathbf{r}) = \frac{1}{2\pi} \iint_S \left[ \frac{\Omega - 1}{|\mathbf{r}_s - \mathbf{r}|} - (\Psi + \phi_-) \mathbf{n} \cdot \frac{\mathbf{r}_s - \mathbf{r}}{|\mathbf{r}_s - \mathbf{r}|^3} \right] dS(\mathbf{r}_s), \tag{48}$$

$$\frac{\partial \Omega}{\partial \tau_s} = -V_s \nabla_s^2 (\phi_- + \Psi), \tag{49}$$

$$\frac{\partial \Psi}{\partial \tau_s} = \frac{1}{2} u_-^2 - V_s^2 + (1 + \Omega) V_s + \frac{\Theta - 1}{2}. \tag{50}$$

Besides, velocity just ahead of the flame front is determined from Eq. (37) as

$$\mathbf{u}_-(\mathbf{r}) = \frac{1}{4\pi} \iint_S \left[ (1 - V_s) \frac{\mathbf{r}_s - \mathbf{r}}{|\mathbf{r}_s - \mathbf{r}|^3} - \phi_- \frac{\partial}{\partial \mathbf{n}} \left( \frac{\mathbf{r}_s - \mathbf{r}}{|\mathbf{r}_s - \mathbf{r}|^3} \right) \right] dS(\mathbf{r}_s). \tag{51}$$

Then the flame front velocity may be calculated as $V_s = u_{n-} - 1$, Eq. (8), using Eq. (51). The system of Eqs. (47) - (51) contains only values and variables at the flame front, thus, reducing the 3D problem Eqs. (1), (2), (8) − (11) in the bulk of the gas flow to a 2D problem of flame dynamics as a discontinuity surface.

The system of Eqs. (47) - (51) holds for any thermal expansion of the burning matter including the realistic expansion factors $\Theta = 5 - 10$, when corrugated flame shape generates vorticity behind the flame front. It is interesting to compare the above equations to the Frankel equation (Frankel, 1990) obtained under the assumption of a potential flow in the burnt matter. As is known, the flow of the burnt gas may be treated as potential in the case of small thermal expansion $\Theta - 1 \ll 1$ (Sivashinsky, 1977). In that case Eq. (49) takes the form of the Laplace equation on a closed surface $\nabla_s^2 (\phi_- + \Psi) = 0$, which has the only solution $\phi_- + \Psi = const$. Since potential of a double layer with constant density is also constant, then in agreement with (Frankel, 1990) Eq. (48) reduces to the velocity potential of a single layer

$$\phi_-(\mathbf{r}) = -\frac{\Theta - 1}{4\pi} \iint_S \frac{dS(\mathbf{r}_s)}{|\mathbf{r}_s - \mathbf{r}|} + const. \tag{52}$$

Then after calculations similar to (Frankel, 1990) we come to the Frankel equation

$$V_s - 1 = -u_{n-} = \frac{\Theta - 1}{2} \left( 1 - \frac{1}{2\pi} \iint_S \frac{\partial}{\partial \mathbf{n}} \frac{1}{|\mathbf{r}_s - \mathbf{r}|} dS(\mathbf{r}_s) \right). \tag{53}$$

Equation (53) may be also reduced to the Sivashinsky equation (Sivashinsky, 1977) in the case of weak nonlinearity, see (Frankel, 1990).

An interesting feature of Eqs. (47) − (51) is that the system obtained involves indirectly the second order derivative in time contrary to the Sivashinsky and Frankel equations, which are only of the first order. The difference between the equations of the second and first order in time may be crucial in description of such phenomena as "tulip flames" (Dold and Joulin, 1995) and flame-shock interactions (Bychkov, 1998; Travnikov et al., 1999).

**5. Equations at a flame front of finite thickness**

In the previous section we have considered equations at an infinitely thin flame front. However, according to the DL theory, the instability growth rate of small perturbations at a flame front is infinitely large if the perturbation wavelength is not limited from below by the cut off wavelength $\Lambda_c = \lambda_c / R$ (proportional to the finite flame thickness). In order to describe thermal stabilization of the DL instability one has to take into account finite flame thickness in the conservation laws Eqs. (5) − (7) (Matalon and Matkowsky, 1982). Rigorous consideration of the finite flame thickness in Eqs. (5) − (7) requires rather long calculations and will be presented elsewhere. In the present section we demonstrate how the effects of thermal stabilization may be taken into account in the system (47) − (51) in a simplified way similar to the classical Markstein approach (Markstein, 1964). For that purpose we can notice that development of the DL instability at both linear and nonlinear stages involve only one parameter of length dimension, namely, the cut off wavelength $\lambda_c$ (Matalon Matkowsky, 1982; Bychkov et al. 1996, Bychkov 1998; Bychkov et al. 1999). Similar to the Markstein approximation we take Eq. (8) in the form

$$u_{n-} + V_s = 1 - \Lambda Y ,\tag{54}$$

where $Y$ is stretch of the flame front (relative increase of the elementary surface area $\Delta$ at the flame front per unit time $Y \equiv \Delta^{-1} d\Delta / d\tau_s$ (Matalon and Matkowsky, 1983)) and $\Lambda$ is a coefficient characterizing thermal stabilization of the DL instability. Other conservation laws Eqs. (9) − (11) are considered without any change. We are going to find the relation between $\Lambda$ and $\Lambda_c$ and use $\Lambda_c$ instead of $\Lambda$. In the linear case of a slightly perturbed flame front propagating in a tube, stretch may be calculated as (Matalon and Matkowsky, 1982)

$$Y = \nabla_\perp \cdot \mathbf{w}_- + \nabla_\perp^2 f ,\tag{55}$$

and one has to replace the last equation of (25) by

$$\frac{\partial f}{\partial \tau} - \tilde{\upsilon}_- = \Lambda \left( \nabla_\perp \cdot \mathbf{w}_- + \nabla_\perp^2 f \right).\tag{56}$$

Then the system (16) − (18), (22) − (25), (56) reduces to the dispersion equation

$$(\Theta + 1)\frac{\partial^2 f}{\partial \tau^2} + 2\Theta\left[1 + \Lambda \frac{\Theta+1}{2\Theta}\hat{\Phi}\right]\hat{\Phi}\frac{\partial f}{\partial \tau} - \Theta(\Theta-1)\left[1 - \Lambda\frac{\Theta+1}{\Theta-1}\hat{\Phi}\right]\hat{\Phi}^2 f = 0 .\tag{57}$$

As we can see from Eq. (57), the DL instability is stabilized at the perturbation wave number $k_c$ satisfying

$$1 - \Lambda k_c \frac{\Theta + 1}{\Theta - 1} = 0, \tag{58}$$

which corresponds to the cut off wavelength $\Lambda_c = 2\pi / k_c$, that is

$$\Lambda = \frac{1}{2\pi} \frac{\Theta - 1}{\Theta + 1} \Lambda_c. \tag{59}$$

Thus Eq. (54) takes the form

$$u_{n-} + V_s = 1 - \frac{1}{2\pi} \frac{\Theta - 1}{\Theta + 1} \Lambda_c Y, \tag{60}$$

where $\Lambda_c = \lambda_c / R$ is the scaled cut off wavelength of the DL instability (the dimensional value of the cut off wavelength $\lambda_c = \Lambda_c R$ may be measured experimentally (Clanet and Searby, 1998)).

Taking into account the stretch effects we find the boundary conditions for the velocity potential

$$\frac{\partial \phi_-}{\partial \mathbf{n}} = u_{n-} = 1 - V_s - \Lambda Y, \tag{61}$$

$$\frac{\partial \phi_+}{\partial \mathbf{n}} = u_{pn+} = \Theta - V_s - \Theta \Lambda Y - u_{vn+}, \tag{62}$$

and the solution to the Laplace equation at the flame front

$$\phi_-(\mathbf{r}) = \frac{1}{2\pi} \iint_S \left[ \frac{1 - V_s - \Lambda Y}{|\mathbf{r}_s - \mathbf{r}|} + \phi_- \mathbf{n} \cdot \frac{\mathbf{r}_s - \mathbf{r}}{|\mathbf{r}_s - \mathbf{r}|^3} \right] dS(\mathbf{r}_s), \tag{63}$$

$$\phi_+(\mathbf{r}) = -\frac{1}{2\pi} \iint_S \left[ \frac{\Theta - V_s - \Theta \Lambda Y - u_{vn+}}{|\mathbf{r}_s - \mathbf{r}|} + \phi_+ \mathbf{n} \cdot \frac{\mathbf{r}_s - \mathbf{r}}{|\mathbf{r}_s - \mathbf{r}|^3} \right] dS(\mathbf{r}_s). \tag{64}$$

The relation (43) is not affected by the finite flame stretch, but Eq. (45) reduces to

$$\frac{\partial}{\partial \tau_s} (\phi_+ - \Theta \phi_-) = \frac{\Theta - 1}{2} u_-^2 - (\Theta - 1) V_s^2 + (\Theta - 1) V_s +$$

$$V_s u_{vn+} - (\Theta - 1)(V_s + \Theta - 1) \Lambda Y + \frac{(\Theta - 1)^2}{2} \left[ 1 + (\Lambda Y)^2 \right]. \tag{65}$$

Then the final set of equations at the flame front takes the form

$$\phi_-(\mathbf{r}) = \frac{1}{2\pi} \iint_S \left[ \frac{1}{|\mathbf{r}_s - \mathbf{r}|} \left( 1 - V_s - \frac{\Theta - 1}{\Theta + 1} \frac{\Lambda_c Y}{2\pi} \right) + \phi_- \mathbf{n} \cdot \frac{\mathbf{r}_s - \mathbf{r}}{|\mathbf{r}_s - \mathbf{r}|^3} \right] dS(\mathbf{r}_s), \tag{66}$$

$$\Psi(\mathbf{r}) + \frac{\Theta+1}{\Theta-1}\phi_-(\mathbf{r}) =$$

$$\frac{1}{2\pi}\iint_S\left[\frac{1}{|\mathbf{r}_s-\mathbf{r}|}\left(\Omega-1+\frac{\Theta-1}{\Theta+1}\frac{\Lambda_c Y}{2\pi}\right)-(\Psi+\phi_-)\mathbf{n}\cdot\frac{\mathbf{r}_s-\mathbf{r}}{|\mathbf{r}_s-\mathbf{r}|^3}\right]dS(\mathbf{r}_s), \qquad (67)$$

$$\frac{\partial\Omega}{\partial\tau_s} = -V_s\nabla_s^2(\phi_-+\Psi), \qquad (68)$$

$$\frac{\partial\Psi}{\partial\tau_s} = \frac{1}{2}u_-^2 - V_s^2 + (1+\Omega)V_s - \frac{\Theta-1}{\Theta+1}\frac{\Lambda_c Y}{2\pi}(V_s+\Theta-1)+$$

$$\frac{\Theta-1}{2}\left[1+\left(\frac{\Theta-1}{\Theta+1}\frac{\Lambda_c Y}{2\pi}\right)^2\right]. \qquad (69)$$

The scaled cut off wavelength of the DL instability $\Lambda_c$ is the only parameter of length dimension involved into Eqs. (66) − (69), and, obviously, $\Lambda_c$ is the smallest length scale that has to be resolved in numerical solution of the above equations.

## 6. Equations at a turbulent flame front

In this section we will show how external turbulence may be included into Eqs. (66) − (69) using the approximation of small vorticity both upstream and downstream the flame front. Small effects of vorticity in the flamelet regime of turbulent burning has been discussed recently in (Denet, 1997; Peters et al., 2000), where vorticity has been neglected completely. Approximation of the present paper is much less restrictive, though, of course, the accuracy of such approximation is considerably lower for turbulent flames in comparison with the laminar ones because the vorticity effects are obviously stronger for turbulent flames. Still, the present approximation is consistent with the Taylor hypothesis of "stationary" turbulence (Williams, 1985), which follows from Eq. (28) applied to the turbulent flow of the fuel mixture. The Taylor hypothesis has not been proven rigorously, but it was used in the majority of papers devoted to turbulent burning in the flamelet regime, see, for example, (Yakhot, 1988; Denet, 1997; Kagan and Sivashinsky, 2000). Recent investigation of flame dynamics in a flow with temporal pulsations of external turbulent velocity (Bychkov and Denet, 2002) has demonstrated, that the Taylor hypothesis does provide a good model for the flamelet regime of burning.

In the case of turbulent flames the upstream flow in the fuel mixture contains both a potential mode and a turbulent mode of vorticity drift, and the boundary conditions for the velocity potential at the flame front is

$$\frac{\partial \phi_-}{\partial \mathbf{n}} = u_{pn-} = u_{n-} - u_{en-} = 1 - V_s - u_{en-} - \Lambda Y, \qquad (70)$$

where $u_{en-}$ is the normal component of the external turbulent velocity at the flame front, and the respective solution to the Laplace equation ahead of the front is

$$\phi_-(\mathbf{r}) = \frac{1}{2\pi} \iint_S \left[ \frac{1}{|\mathbf{r}_s - \mathbf{r}|} \left( 1 - V_s - u_{en-} - \frac{\Theta - 1}{\Theta + 1} \frac{\Lambda_c Y}{2\pi} \right) + \phi_- \mathbf{n} \cdot \frac{\mathbf{r}_s - \mathbf{r}}{|\mathbf{r}_s - \mathbf{r}|^3} \right] dS(\mathbf{r}_s), \qquad (71)$$

with the velocity in the fuel mixture at the flame front

$$\mathbf{u}_- = \mathbf{u}_e + \frac{1}{4\pi} \iint_S \left[ \left( 1 - V_s - u_{en-} - \frac{\Theta - 1}{\Theta + 1} \frac{\Lambda_c Y}{2\pi} \right) \frac{\mathbf{r}_s - \mathbf{r}}{|\mathbf{r}_s - \mathbf{r}|^3} - \phi_- \frac{\partial}{\partial \mathbf{n}} \left( \frac{\mathbf{r}_s - \mathbf{r}}{|\mathbf{r}_s - \mathbf{r}|^3} \right) \right] dS(\mathbf{r}_s). \quad (72)$$

The equation relating the vorticity modes and the potential modes at the flame front takes the form

$$\frac{\partial}{\partial \tau_s} \left( u_{vn+} - u_{en-} \right) = V_s \nabla_s^2 \left( \phi_- - \phi_+ \right), \qquad (73)$$

while the equation coupling two potential modes upstream and downstream the flame front coincides with Eq. (65). Then the final system of equations at a turbulent flame front is

$$\phi_-(\mathbf{r}) = \frac{1}{2\pi} \iint_S \left[ \frac{1}{|\mathbf{r}_s - \mathbf{r}|} \left( 1 - V_s - u_{en-} - \frac{\Theta - 1}{\Theta + 1} \frac{\Lambda_c Y}{2\pi} \right) + \phi_- \mathbf{n} \cdot \frac{\mathbf{r}_s - \mathbf{r}}{|\mathbf{r}_s - \mathbf{r}|^3} \right] dS(\mathbf{r}_s), \qquad (74)$$

$$\Psi(\mathbf{r}) + \frac{\Theta + 1}{\Theta - 1} \phi_-(\mathbf{r}) =$$

$$\frac{1}{2\pi} \iint_S \left[ \frac{1}{|\mathbf{r}_s - \mathbf{r}|} \left( \Omega - \frac{u_{en-}}{\Theta - 1} - 1 + \frac{\Theta - 1}{\Theta + 1} \frac{\Lambda_c Y}{2\pi} \right) - (\Psi + \phi_-) \mathbf{n} \cdot \frac{\mathbf{r}_s - \mathbf{r}}{|\mathbf{r}_s - \mathbf{r}|^3} \right] dS(\mathbf{r}_s), \quad (75)$$

$$\frac{\partial}{\partial \tau_s} \left( \Omega - \frac{u_{en-}}{\Theta - 1} \right) = -V_s \nabla_s^2 (\phi_- + \Psi), \qquad (76)$$

$$\frac{\partial \Psi}{\partial \tau_s} = \frac{1}{2} u_-^2 - V_s^2 + \left( 1 - \frac{\Theta u_{en-}}{\Theta - 1} + \Omega \right) V_s - \frac{\Theta - 1}{\Theta + 1} \frac{\Lambda_c Y}{2\pi} (V_s + \Theta - 1) +$$

$$\frac{\Theta - 1}{2} \left[ 1 + \left( \frac{\Theta - 1}{\Theta + 1} \frac{\Lambda_c Y}{2\pi} \right)^2 \right]. \qquad \mathbf{(77)}$$

Of course, in reality turbulent velocity field $\mathbf{u}_e$ at the flame front does not coincide with the velocity field far ahead of the flame front. However, present knowledge about the initial "free" turbulence induced in gas turbines and car engines is very limited. Though the standard assumption about the external velocity field used in numerical simulations (Denet, 1997; Kagan and Sivashinsky, 2000; Peters et al., 2000; Bychkov and Denet, 2002) is the assumption of an isotropic Kolmogorov turbulence, one cannot say for sure that such turbulence takes place in combustion experiments (Abdel-Gayed et al., 1988; Aldredge et al., 1998; Kobayashi et al., 1998) and in industrial energy production devices. Therefore, instead of making assumptions about turbulence far ahead of the flame front, at present one can make assumptions directly about the turbulent velocity field at the flame front.

## 7. Equations at a flame front in a two-dimensional geometry

Though equations (74) – (77) do reduce the hydrodynamic problem in the bulk of the gas flow to a set of equations at the flame front, the resulting system is still rather complicated for numerical solution. By this reason it is natural to expect that the first modelling of Eqs. (74) – (77) will be performed in the 2D geometry rather than the 3D one. In this section we present the 2D version the system (74) – (77), for which the flame front is a curve instead of a surface. We start with the infinitely thin laminar flame. In that case the integral expression for the Green solution to the Laplace equation (37) takes the form

$$\gamma \phi(\mathbf{r}) = \int_S \left[ \frac{\partial \phi(\mathbf{r}_s)}{\partial \mathbf{n}_{out}} \ln|\mathbf{r}_s - \mathbf{r}| - \phi(\mathbf{r}_s)\mathbf{n}_{out} \cdot \frac{\mathbf{r}_s - \mathbf{r}}{|\mathbf{r}_s - \mathbf{r}|^2} \right] dS(\mathbf{r}_s), \tag{78}$$

where the curve $S$ is the boundary of a 2D-domain $G$. The factor $\gamma$ is zero $\gamma = 0$ if $\mathbf{r}$ is outside $G$, $\gamma = \pi$ if $\mathbf{r}$ belongs to the curve $S$, and $\gamma = 2\pi$ if $\mathbf{r}$ is inside $G$. Therefore, at the flame front we have the following expressions for the velocity potentials

$$\phi_+(\mathbf{r}) = -\frac{1}{\pi} \int_S \left[ (\Theta - V_s - u_{vn+})\ln|\mathbf{r}_s - \mathbf{r}| - \phi_+\mathbf{n} \cdot \frac{\mathbf{r}_s - \mathbf{r}}{|\mathbf{r}_s - \mathbf{r}|^2} \right] dS(\mathbf{r}_s), \tag{79}$$

$$\phi_-(\mathbf{r}) = \frac{1}{\pi} \int_S \left[ (1 - V_s)\ln|\mathbf{r}_s - \mathbf{r}| - \phi_-\mathbf{n} \cdot \frac{\mathbf{r}_s - \mathbf{r}}{|\mathbf{r}_s - \mathbf{r}|^2} \right] dS(\mathbf{r}_s). \tag{80}$$

The equation (80) is the 2D counterpart of Eq. (47). The counterpart of Eq. (48) follows from Eqs. (79), (80)

$$\Psi(\mathbf{r}) + \frac{\Theta+1}{\Theta-1}\phi_-(\mathbf{r}) = \frac{1}{\pi}\int\limits_S\left[(\Omega-1)\ln|\mathbf{r}_s - \mathbf{r}| + (\Psi+\phi_-)\mathbf{n}\cdot\frac{\mathbf{r}_s - \mathbf{r}}{|\mathbf{r}_s - \mathbf{r}|^2}\right]dS(\mathbf{r}_s). \tag{81}$$

Eqs. (49), (50) remain the same in the 2D geometry as they were in the 3D one

$$\frac{\partial\Omega}{\partial\tau_s} = -V_s\nabla_s^2(\phi_- + \Psi), \tag{82}$$

$$\frac{\partial\Psi}{\partial\tau_s} = \frac{1}{2}u_-^2 - V_s^2 + (1+\Omega)V_s + \frac{\Theta-1}{2}, \tag{83}$$

and the velocity just ahead of the flame front may be found from Eq. (78)

$$\mathbf{u}_-(\mathbf{r}) = \frac{1}{2\pi}\int\limits_S\left[-(1-V_s)\frac{\mathbf{r}_s - \mathbf{r}}{|\mathbf{r}_s - \mathbf{r}|^2} + \phi_-\frac{\partial}{\partial\mathbf{n}}\left(\frac{\mathbf{r}_s - \mathbf{r}}{|\mathbf{r}_s - \mathbf{r}|^2}\right)\right]dS(\mathbf{r}_s). \tag{84}$$

The set of equations (80) - (84) presents a coordinate-free description of 2D laminar flames with infinitely small flame thickness. Similar to Sec. 5 and 6 we may take into account stretch effects (produced by finite flame thickness) and external turbulence. Then the system (80) - (84) takes the form

$$\phi_-(\mathbf{r}) = \frac{1}{\pi}\int\limits_S\left[\left(1 - V_s - u_{en-} - \frac{\Theta-1}{\Theta+1}\frac{\Lambda_c Y}{2\pi}\right)\ln|\mathbf{r}_s - \mathbf{r}| - \phi_-\mathbf{n}\cdot\frac{\mathbf{r}_s - \mathbf{r}}{|\mathbf{r}_s - \mathbf{r}|^2}\right]dS(\mathbf{r}_s), \tag{85}$$

$$\Psi(\mathbf{r}) + \frac{\Theta+1}{\Theta-1}\phi_-(\mathbf{r}) =$$

$$\frac{1}{\pi}\int\limits_S\left[\left(\Omega - \frac{u_{en-}}{\Theta-1} - 1 + \frac{\Theta-1}{\Theta+1}\frac{\Lambda_c Y}{2\pi}\right)\ln|\mathbf{r}_s - \mathbf{r}| + (\Psi+\phi_-)\mathbf{n}\cdot\frac{\mathbf{r}_s - \mathbf{r}}{|\mathbf{r}_s - \mathbf{r}|^2}\right]dS(\mathbf{r}_s), \tag{86}$$

$$\frac{\partial}{\partial\tau_s}\left(\Omega - \frac{u_{en-}}{\Theta-1}\right) = -V_s\nabla_s^2(\phi_- + \Psi), \tag{87}$$

$$\frac{\partial\Psi}{\partial\tau_s} = \frac{1}{2}u_-^2 - V_s^2 + \left(1 - \frac{\Theta u_{en-}}{\Theta-1} + \Omega\right)V_s - \frac{\Theta-1}{\Theta+1}\frac{\Lambda_c Y}{2\pi}(V_s + \Theta - 1) +$$

$$\frac{\Theta-1}{2}\left[1 + \left(\frac{\Theta-1}{\Theta+1}\frac{\Lambda_c Y}{2\pi}\right)^2\right], \tag{88}$$

$$\mathbf{u}_-(\mathbf{r}) = \mathbf{u}_e - \frac{1}{2\pi}\int\limits_S \left[ \left(1 - V_s - u_{en-} - \frac{\Theta - 1}{\Theta + 1}\frac{\Lambda_c Y}{2\pi}\right) \frac{\mathbf{r}_s - \mathbf{r}}{|\mathbf{r}_s - \mathbf{r}|^2} - \phi_- \frac{\partial}{\partial \mathbf{n}}\left(\frac{\mathbf{r}_s - \mathbf{r}}{|\mathbf{r}_s - \mathbf{r}|^2}\right) \right] dS(\mathbf{r}_s). \qquad (89)$$

## 8. Summary

In the present paper we have reduced the whole system of hydrodynamic equations in the bulk of the gas flow ahead of a corrugated flame front and behind the front to a set of equations at the flame front, see Eqs. (47) – (50) for a laminar flame front of zero thickness, Eqs. (66) – (69) for a laminar flame front of finite thickness, and Eqs. (74) – (77) for a flame front in an external turbulent flow. The derived equations may provide considerable gain in numerical simulations of laminar corrugated flames and turbulent burning in the flamelet regime. First, the derived equations reduce the dimension of the problem by one, since a 3D problem of the gas flow is replaced by a 2D problem of the flame front dynamics considered as a geometrical surface. Second, the smallest length scale involved in the equations, which has to be resolved in the numerical simulations, is the cut off wavelength of the DL instability $\lambda_c$. This length scale is almost three orders of magnitude larger that the thickness of the reaction zone, which has to be resolved in the direct numerical simulations. Indeed, the thickness of the reaction zone with realistically large activation energy of the chemical reactions is usually about 0.1 of the flame thickness, see (Bychkov et al., 1996; Kadowaki, 1999; Travnikov et al., 2000). On the other hand, the cut off wavelength $\lambda_c$ typically exceeds the flame thickness by a factor of $40 - 50$ (Pelce and Clavin, 1982; Serby and Rochwerger, 1992). Thus the cut off wavelength $\lambda_c$ is larger than the thickness of the reaction zone by a factor of $400 - 500$.

     With all these advantages of the obtained equations (74) – (77) one may hope to model turbulent burning in realistic energy production devices, for which the characteristic length scale of the hydrodynamic flow $(10 \text{ cm} - 1 \text{ m})$ exceeds the thickness of the reaction zone by $5 - 6$ orders of magnitude making these flows far beyond the reach of direct numerical simulations. Still, as a next step of the research, the equation obtained have to be validated by comparing the numerical results of the model to experiments and direct numerical simulations. This will be the subject of the future work.

## Acknowledgments

This work was supported by the Swedish Research Council (VR).

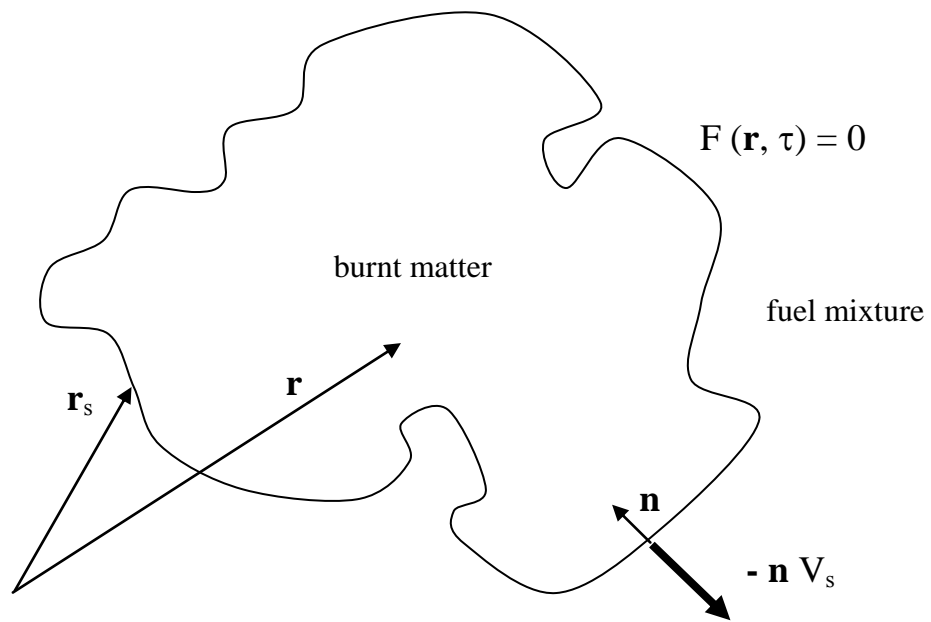

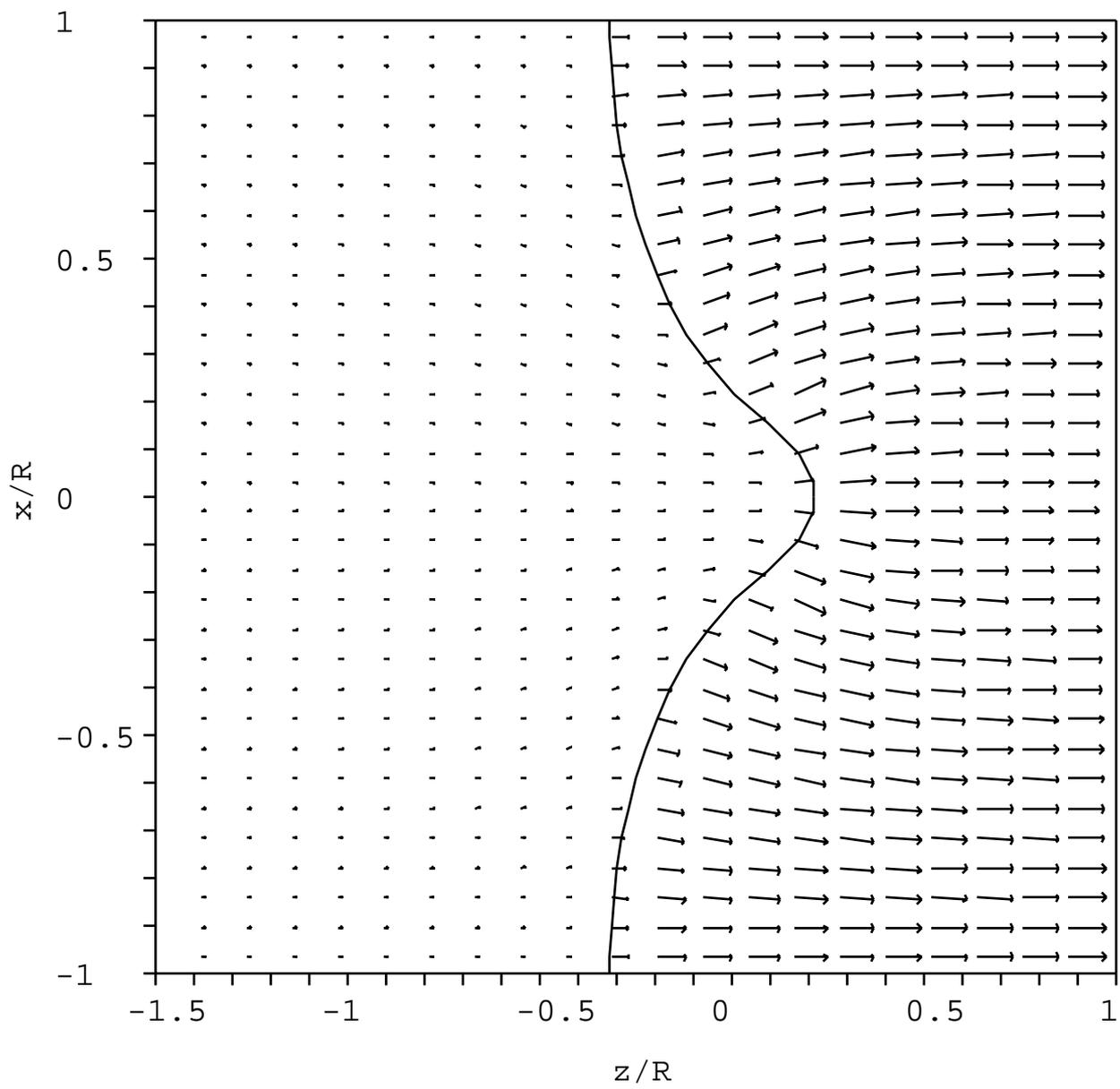